\documentclass[]{aastex62}
\usepackage{longtable}
\usepackage{amsmath}
\usepackage{subfigure}
\usepackage{appendix}

\graphicspath{{./}{figures/}}

\shorttitle{UHDM searches}
\shortauthors{Tak et al.}

\begin{document}

\reportnum{\footnotesize CERN-TH-2022-139}
\reportnum{\footnotesize DESY-22-138}

\title{Current and future $\gamma$-ray searches for dark-matter annihilation beyond the unitarity limit}
\correspondingauthor{Donggeun Tak (\href{mailto:donggeun.tak@gmail.com}{donggeun.tak@gmail.com})}
\author{Donggeun Tak}
\affiliation{Deutsches Elektronen-Synchrotron DESY, Platanenallee 6, 15738, Zeuthen, Germany}

\author{Matthew Baumgart}
\affiliation{Department of Physics, Arizona State University, Tempe, AZ 85287, USA}

\author{Nicholas L. Rodd}
\affiliation{Theoretical Physics Department, CERN, 1 Esplanade des Particules, CH-1211 Geneva 23, Switzerland}

\author{Elisa Pueschel}
\affiliation{Deutsches Elektronen-Synchrotron DESY, Platanenallee 6, 15738, Zeuthen, Germany}

\begin{abstract}
For decades, searches for electroweak-scale dark matter (DM) have been performed without a definitive detection.
This lack of success may hint that DM searches have focused on the wrong mass range.
A proposed candidate beyond the canonical parameter space is ultra-heavy DM (UHDM).
In this work, we consider indirect UHDM annihilation searches for masses between 30 TeV and 30 PeV---extending well beyond the unitarity limit at $\sim$100 TeV---and discuss the basic requirements for DM models in this regime.
We explore the feasibility of detecting the annihilation signature, and the expected reach for UHDM with current and future Very-High-Energy (VHE; $>$\,100\,GeV) $\gamma$-ray observatories.
Specifically, we focus on three reference instruments: two Imaging Atmospheric Cherenkov Telescope arrays, modeled on VERITAS and CTA-North, and one Extended Air Shower array, motivated by HAWC.
With reasonable assumptions on the instrument response functions and background rate, we find a set of UHDM parameters (mass and cross section) for which a $\gamma$-ray signature can be detected by the aforementioned observatories.  We further compute the expected upper limits for each experiment.
With realistic exposure times, the three instruments can probe DM across a wide mass range.  At the lower end, it can still have a point-like cross section, while at higher masses the DM could have a geometric cross section, indicative of compositeness.
\end{abstract}

\keywords{Dark Matter, Ultra-heavy Dark Matter}

\section{Introduction} \label{sec:intro}

Dark matter (DM) is an unrevealed component of the matter in the Universe whose existence is widely supported by a broad set of observations \citep{Bertone2018a}.
For decades, many theoretical candidates have been considered for particle DM, of which two representative examples are ultralight axions ($M_{\chi}$ $\ll$ 1\,eV) and weakly interacting massive particles (WIMPs; $M_{\chi} \sim {\cal O}({\rm GeV}-{\rm TeV})$).
Both candidates have been hunted for with state-of-the-art experiments and observatories, and although these searches will continue to achieve important milestones---for example the long sought-after Higgsino may soon be within reach~\citep{Rinchiuso:2020skh,Dessert:2022evk}---so far the program has been unsuccessful \citep[for the latest reviews, see e.g.,][]{Gaskins2016, Boveia2018, Tao2020}.

The longstanding lack of a DM signal detection has driven theorists to look for DM candidates beyond the conventional parameter space.
One such candidate is ultra-heavy DM (UHDM; 10\,TeV  $\lesssim M_{\chi} \lesssim m_{\rm pl} \approx 10^{19}$\,GeV).
Depending on the cosmological scenario and beyond the Standard Model (SM) theory that predicts the UHDM, the abundance and properties can vary \citep[for a broad outline, see][]{snowmass2022}; e.g., WIMPzilla \citep{Kolb1999} and Gluequark DM \citep{Contino2019}.
In addition to unexplored UHDM candidates, there are models that extend the WIMP mass range beyond $\sim$10\,TeV \citep[e.g.,][]{Harling2014, Baldes2017, Cirelli2019, Bhatia2021}.
Yet there exists a general upper limit on the WIMP mass, known as the unitarity limit, which requires $M_{\chi} \lesssim 194$\,TeV \citep{Griest:1989wd,Smirnov:2019ngs}.
This bound arises as the standard WIMP paradigm is associated with a thermal relic cosmology.
In this scenario, in the early Universe, the DM and SM particles are in thermal equilibrium.
As the Universe expands and cools, the DM departs from equilibrium and its abundance is rapidly depleted by annihilations, until the expansion eventually shuts this process off and the relic abundance freezes out.
The key parameter in this scenario is the DM annihilation cross section, which for point-like particles going to Standard Model (SM) states must scale as $M_{\chi}^{-2}$ by dimensional analysis.
As the mass increases, the cross section generally decreases. If it becomes too small, then the DM will be insufficiently depleted by the time it freezes out, and too much DM will remain to be consistent with the observed cosmological density.
Ultimately, as unitarity dictates that the cross section cannot be made arbitrarily large, this constraint translates into the stated upper bound on the DM mass.

While there is an attractive simplicity to the thermal-relic cosmology so described, as soon as we allow even minimal departures from it, the unitarity bound can be violated, allowing for the possibility that DM with even higher masses could be annihilating in the present day Universe.
For example, instead of annihilating directly to SM states, the DM could produce a metastable dark state which itself decays to the SM.
As shown by \cite{Berlin2016}, if this dark state lives long enough to dominate the energy density of the Universe, its decays to the SM will then dilute the DM density, avoiding the overproduction otherwise associated with heavy thermal DM, and allowing masses up to 100\,PeV to be obtained.
PeV-scale thermal DM can also be achieved if the DM is a composite state, rather than a point-like particle.
Exactly such a scenario was considered by \cite{Harigaya:2016nlg}, where DM with a large radius arose from a model of a strongly coupled confining theory in the dark sector.  The lightest baryon in the theory plays the role of DM, which annihilates through a portal coupling to eventually produce SM states.
Such a scenario can evade the unitarity bound as the annihilation cross section is no longer guaranteed to scale as $M_{\chi}^{-2}$; it can instead now be determined by the geometric size of the composite DM.
Indeed, we will see that such composite DM scenarios are broadly the models that can be probed using the observational strategies considered in this work.

The self-annihilations which play a role in setting DM abundance in the early Universe can also be active today, producing an observable flux of stable SM particles such as $e^{\pm}$, $\nu_{e, \mu, \tau}$, and $\gamma$-rays, as well as unstable quarks, leptons, and bosons whose interaction processes can produce secondary $\gamma$-rays.
The full energy spectrum at production can be estimated with Monte Carlo (MC) simulations of the underlying particle physics.
For this purpose, {\tt PYTHIA} is the most widely used program, providing an accurate prompt DM spectrum up to ${\cal O}(10)$ TeV \citep{pythia}, and is a central ingredient in the widely used PPPC4DMID~\citep{Cirelli:2010xx,Ciafaloni:2010ti}.
However, {\tt PYTHIA} is not appropriate for studying UHDM in general, as it omits many of the interactions in the full unbroken SM that become important as the UHDM mass becomes much larger than the electroweak scale.
An alternative approach was introduced in \cite{HDM}, which computed the prompt DM spectrum from 1 TeV up to the Planck scale, the so-called {\tt HDMSpectrum}.\footnote{The results are publicly available at \url{https://github.com/nickrodd/HDMSpectra}.}
To do so, the authors of that work mapped the calculation of the DM spectrum to the computation of fragmentation functions, which can then be computed with DGLAP evolution in a manner that includes all relevant SM interactions, providing a better characterization of the prompt UHDM spectrum \citep[see][for a discussion of earlier approaches to compute DM spectra]{HDM}.

When $\gamma$-rays are produced from DM annihilation throughout the Universe, they can propagate to the Earth and be detected.
After considering the propagation effects,\footnote{For DM searches with galaxies in the Local Group, any galactic absorption by the starlight, infrared photons, and/or cosmic microwave background can be ignored due to its relatively small contribution \citep[$<$20\% at $\mathcal{O}$(100) TeV;][]{Esmaili2015}.
We note that while the UHDM mass range considered extends to 30 PeV, detected photons with energies above 100 TeV are not considered.}
the $\gamma$-ray flux at the Earth from DM annihilation can be described as
\begin{equation}\label{eq:dm_flux}
    \frac{dF(E, \hat{n})}{dE d\Omega} = \frac{\langle\sigma v\rangle}{8\pi M_{\chi}^2}\frac{dN_{\gamma}(E)}{dE}\int_{\rm los}dl\, \rho^2(l\hat{n}),
\end{equation}
where $\langle\sigma v\rangle$ is the velocity-averaged annihilation cross section. 
The prompt energy spectrum, $dN_{\gamma}(E)/dE$, depends on the DM annihilation channel and is determined from the HDM spectrum; $\rho(l\hat{n})$ is the DM density along the line of sight (los).
Even though the DM annihilation process can occur anywhere that DM is present, the DM signature from DM-rich regions will be brighter.
For instance, dwarf spheroidal galaxies (dSphs) in the Local Group are one of the best targets for DM study because of their high mass-to-light ratio (implying high DM density; e.g., $M/L \sim 3400 M_{\sun}/L_{\sun}$ for Segue 1; \citealp{Simon2011}), close proximity, and absence of bright nearby background sources.

The $\gamma$-rays that could be arriving at Earth from DM annihilations would be detectable with $\gamma$-ray space telescopes and ground-based observatories, enabling indirect searches for DM.
The self-annihilation of UHDM can produce $\gamma$-rays from around a TeV to above a PeV, containing the energy band in which the ground-based $\gamma$-ray observatories have better sensitivity than space-based instruments.
There are two classes of ground-based Very-High-Energy (VHE; $>$100 GeV) $\gamma$-ray observatories: Imaging Atmospheric Cherenkov Telescope arrays (IACTs) and Extended Air Shower arrays (EAS). 
IACTs use reflecting dishes and fast cameras (generally using photomultiplier tubes, or PMTs) to reconstruct the Cherenkov light stimulated by the air showers triggered by TeV $\gamma$-rays as they interact with Earth's atmosphere.
Current generation EAS arrays are made of water tanks, where optical detectors (generally PMTs) in each tank directly detect the Cherenkov radition from charged air shower particles. Both types of instrument can reconstruct TeV $\gamma$-rays~\citep{doi:10.1146/annurev-nucl-102014-022036}.
Both have been used for indirect DM searches, with a particular focus on searches for electroweak-scale WIMPs \citep[e.g.,][]{dm_magic, dm_veritas, dm_hess, dm_hawc, dm_magic2, hawc_dm_halo}. 
In addition to those $\gamma$-ray observatories, neutrino observatories have also searched for an indirect DM signal \citep[e.g.,][]{icecube_dm, Albert2022}.

In this paper, we explore the feasibility of detecting a UHDM annihilation signature from dSphs with current and future ground-based VHE $\gamma$-ray observatories.
To this end, we use only publicly available resources.
Also, we compute expected upper limits (ULs) for an UHDM particle with a mass from 30 TeV to 30 PeV, assuming that the UHDM signal is not detected.
We take Segue 1, one of the local classical dSphs, as our benchmark target, because it has been widely used for indirect DM searches, making it possible to place our results in the context of the existing limits at lower masses \citep[e.g.,][]{dm_veritas, dm_magic}. Furthermore, it has good visibility (in terms of zenith angle of observation) for all of the instruments discussed in this work.
We consider three instruments: the Very Energetic Radiation Imaging Telescope Array System (VERITAS; IACT), the Cherenkov Telescope Array (CTA; IACT), and the High-Altitude Water Cherenkov Observatory (HAWC; EAS array).
For VERITAS and HAWC, we do not access the official instrument response functions (IRFs)\footnote{The IRFs describe the mapping between the true and detected flux, primarily consisting of the effective area, point spread function, and energy dispersion matrix, each of which will differ between experiments.} and/or observed background spectra, but rather make reasonable assumptions based on publicly available information, and introduce a VERITAS-like and a HAWC-like instrument.

The remaining discussion is organized as follows.
In Section~\ref{sec:theory}, we present the theoretical motivations for UHDM searches, with a particular focus on the experimentally accessible parameter space.
The data acquisition and processing for each instrument is detailed in Sec.~\ref{sec:data}, with the methods used to calculate the projected sensitivity and ULs for each instrument outlined in Sec.~\ref{sec:method}.
We present our results in Sec.~\ref{sec:result}, and the studies on the systematic and statistical uncertainties are discussed in Sec.~\ref{sec:discussion}.
Our conclusions are reserved for Sec.~\ref{sec:summary}.

\vspace{0.3in}

\section{Theoretical Motivation}\label{sec:theory}

\begin{figure}[t!]
    \centering
    \includegraphics[width=0.45\linewidth]{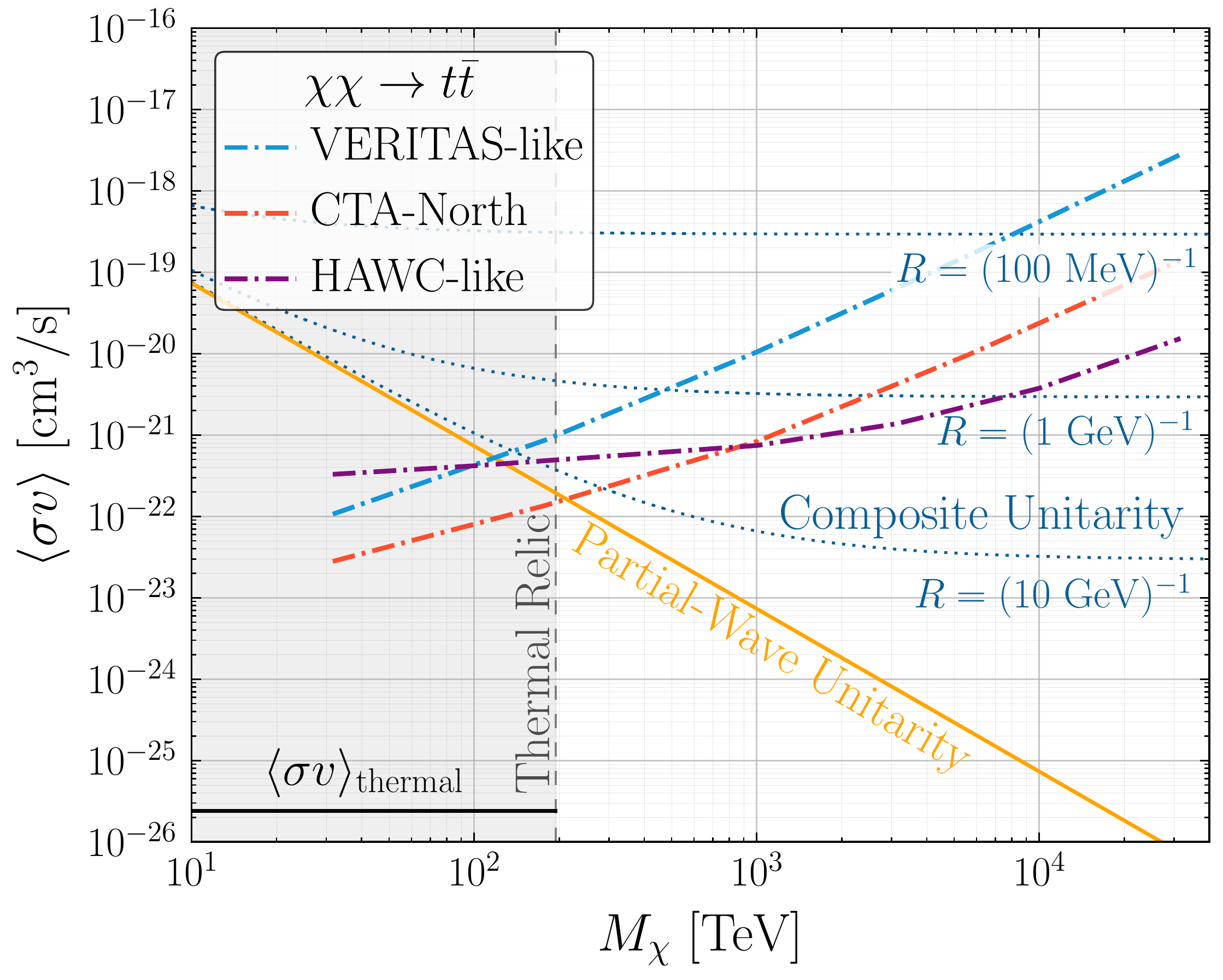}
    \caption{A comparison of our estimated limits for annihilation to $t\bar{t}$ against various theoretical benchmarks.
    The black solid curve refers to the standard thermal-relic cross section ($2.4\times10^{-26}$ cm$^3$/s;~\citealp{Steigman:2012nb}), and the region shaded in gray is the conventional parameter space associated with a point-like thermal relic.
    For Segue 1, the $J=0$ partial-wave unitarity limit on a point-like annihilation cross section is shown in orange---irrespective of the early Universe cosmology, the point-like particles can only annihilate at a rate below this.
    Composite states are not so restricted, however, and can annihilate up to the various composite unitarity bounds.
    For a detailed discussion, see Sec.~\ref{sec:theory}.}
    \label{fig:lim}
\end{figure}

Theoretical arguments for DM have often downplayed the ultraheavy mass regime.
The prejudice against heavier masses arises from the so-called unitarity limit of \cite{Griest:1989wd}, which is based on the following ``bottom-up'' argument.
The naive expectation is that DM annihilation rates for point-like particles will scale as $\langle \sigma v \rangle \sim C/M_{\chi}^2$, where $M_{\chi}$ is the particle mass, and $C$ is a dimensionless parameter.
For a thermal relic, this cross section is what depletes the DM abundance away from its equilibrium value once the temperature of the Universe drops below $M_{\chi}$, and so we expect $\Omega_{\chi} \propto 1/\langle \sigma v \rangle$.
Accordingly, for too-large $M_{\chi}$, DM cannot destroy itself with enough vigor, and the Universe overcloses.
One can boost the size of $C$, but only up to an amount allowed by unitarity.
DM as a {\it simple} self-annihilating thermal relic is only possible for masses up to $\sim$194 TeV \citep{Smirnov:2019ngs}.
We show this upper limit in Fig.~\ref{fig:lim}; 194 TeV is an updated value of the conservative bound from \cite{Griest:1989wd} (those authors used $\Omega_\chi h^2 = 1$, as opposed to the current measurement of $\Omega_\chi h^2 = 0.12$ given by \citealp{Planck:2018vyg}).

To derive $M_{\chi} \lesssim 194~{\rm TeV}$, one assumes that the annihilation rate saturates the unitarity limit ($\langle \sigma v \rangle \propto 1/v$; cf. Eq.~\ref{eq:unitlim} with $J=0$) for the entire relevant history of the DM.
A rate that scales inversely with velocity is typically found only at low velocities and in the presence of a long-range force, as in the celebrated case of Sommerfeld enhancement.
As discussed below, it is difficult to model-build a scenario where the cross section is maximally large, but where the DM continues to behave as a simple elementary particle.
Typically, bound state and compositeness effects will enter in this limit.
For such reasons, in \cite{Griest:1989wd}, the authors felt the above cross-section scaling was overly conservative.
Instead, they assumed that the cross section was dominantly $S$-wave ($\langle \sigma v \rangle \propto v^0$) but with a maximum value still set by unitarity (as given in Eq.~\ref{eq:unitlim}).
Using this, and assuming $\Omega_{\chi} h^2 = 1$, they derived the well-known upper limit of 340 TeV.
Repeating their calculation for $\Omega_{\chi} h^2 = 0.12$, the bound is reduced to $M_{\chi} \lesssim 116~{\rm TeV}$.
Nevertheless, we will adopt the more conservative value of 194 TeV in our results.
It involves the fewest assumptions about the early Universe, but amounts to assuming that DM finds a way to annihilate at the limiting cross-section value throughout the era that set its relic abundance.

The presence of additional structure in either the DM particles themselves or the final states they capture into can weaken even this conservative limit, though.
For example, if capture into bound states is possible, then selection rules can open up annihilation channels into higher partial waves.
The total relic abundance of DM is necessarily set by the sum over all channels, but each partial wave respects the limit from unitarity,
\begin{equation}
\sigma_J \leq \frac{4\pi\, (2J+1)}{M_\chi^2 v_{\rm rel}^2}.
\label{eq:unitlim}
\end{equation}
As discussed in \cite{Bottaro:2021snn}, even for the straightforward scenario of thermal relics that are just multiplets of the electroweak group SU(2)$_L$, this allows DM consistent with unitarity up to $\sim$325 TeV.  
It would seem uncontroversial to analyze the full regime that allows this simple scenario.

To relax the bound farther, as mentioned above, the unitarity limit of roughly one hundred TeV assumes a point-like particle.  This was explicitly recognized in the classic 1990 reference on the matter.
If, however, DM is a composite particle, then the relevant dimensionful scale that sets the annihilation rate can be its geometric size, $R$, which may be much larger than its Compton wavelength $\sim 1/M_{\chi}$.
It is thus possible to realize a thermal-relic scenario for masses $\gg$ 100 TeV ({\it e.g.,} the example of \citealp{Harigaya:2016nlg} discussed above).\footnote{Alternatively, to get to very high masses, one can decouple the DM abundance from its annihilation rate.  In this approach, one forfeits the WIMP-miracle in favor of an alternate cosmological history.
As an example, some other particle could populate the Universe, which ultimately decays to the correct quantity of DM ({\it cf.}~\citealp{Carney:2022gse} for discussion and references).
If DM is non-thermal, then additional structure is needed for detection.  One straightforward possibility is to construct DM that is cosmologically stable, but decays with an observable rate ({\it e.g.}~\citealp{Kolb1999}).}  
For pointing telescopes like VERITAS, HESS, or CTA to have a discovery advantage, one needs a scenario, like compositeness, with non-negligible DM annihilation, since the resulting flux will scale like $\rho^2$.
Bound-state particles with a heavy constituent, whether obtained as thermal relics or by a more complicated cosmology, provide a means to get annihilation rates $\langle \sigma v \rangle \, \gg \, C_{\rm unitary}/M_{\chi}^2$, where $C_{\rm unitary}$ is the largest factor consistent with quantum mechanics in a single partial wave.
One may therefore consider this as a generalization of the ``sum over partial waves'' loophole we first mentioned in the bound-state capture scenario.
As we see in Fig.~\ref{fig:lim}, there is a large region of parameter space beyond the point-like unitarity limit.
Furthermore, we project that the limits from CTA exceed those from HAWC out to several PeV, and are primed for testing these models.

The generic possibility of a geometric cross section for composite particles can be seen with atomic (anti)hydrogen, as pointed out in \cite{Geller:2018biy}, whose arguments we briefly recap.
In a hydrogen-antihydrogen collision, an interaction with a geometric cross section is the ``rearrangement'' reaction, which produces a protonium ($p \bar p$) $+$ positronium ($e^+ e^-$) final state.
Partial-wave by partial-wave, unitarity is naturally respected.
However, summing over all allowed angular momenta gives
\begin{equation}
\sigma \, \sim \, \sum_{J = 0}^{J_{\rm max}}  \sigma_J \, \sim \, \frac{4\pi}{k_i^2} \sum_{J = 0}^{J_{\rm max}} (2J + 1) \,\sim\, \frac{4\pi}{k_i^2} J_{\rm max}^2 \,\sim\, 4\pi R^2,
\label{eq:unitsc}
\end{equation}
where $k_i$ is the initial momentum, $R$ is the size of the particle, and $J_{\rm max}$ is set by angular momentum conservation and the classical value $(k_i \, R)$.\footnote{For this parametric estimate, we are taking $J_{\rm max} \sim L_{\rm max} \sim k_i \, R$.  Strictly, $k_i \, R$ is bounding the orbital angular momentum in the collision.  
Also, Eq.~\eqref{eq:unitsc} assumes a kinetic energy, $E_i = k_i^2/2M_{\chi}$ comparable or larger than the incoming particle's binding energy, $E_b$. 
If $E_i \ll E_b$, then only the $S$-wave will contribute and the cross section $\sigma \sim R/k_i$.
Since this involves just a single partial wave, we therefore cannot use a sum with many terms to exceed the point-particle unitarity limit.}
Importantly, a parametric enhancement in the cross section has been achieved by saturating each partial-wave bound up to $J_{\rm max}$.

Whatever partial-wave protonium is captured into, it will ultimately decay down the spectroscopic ladder until reaching the lowest-allowed-energy state, at which point it annihilates.
For a generic scenario with the dark sector charged under the SM, the entire process of capture, decay, and annihilation is prompt on observational timescales.
An ultraheavy dark-hydrogen thus provides a proof of concept for a ``detection-through-annihilation'' scenario.
The argument for geometric scaling generalizes, though, to include states bound by strong dynamics \citep{Kang:2006yd,Jacoby:2007nw}.
Thus, DM may be more like an ultraheavy $B$-meson \cite[as studied by][]{Geller:2018biy}, or a gluequark \citep[adjoint fermion with color neutralized by cloud of dark gluons;][]{Contino2019}, heavy-light baryon \citep{Harigaya:2016nlg}, {\it etc.}
For a complete scenario, one would necessarily need an explanation for why these heavy-constituent composites came to be the DM with the right abundance.
Nonetheless, the physics behind their ability to annihilate with an effective rate far above the point-particle unitarity limit is straightforward.
Therefore, models with dynamics not-too-different from the SM can realize annihilating particle DM all the way to the Planck scale, and should be tested.

With the above in mind, in Fig.~\ref{fig:lim}, we outline basic theoretical aspects of the parameter space we will consider \citep[\textit{cf.}][]{ANTARES:2022aoa}.
Firstly, we see that the majority of the mass range probed is above the conventional unitarity limit. 
Next, the curve we label by ``Partial-Wave Unitarity'' represents the largest present day annihilation cross section consistent with the same point-particle unitarity constraints that when applied in the early Universe constrains $M_{\chi} \lesssim 194$ TeV.
In particular, we require $\langle \sigma v \rangle \leq 4\pi/(M_{\chi}^2 v_{\rm rel})$, where we take $v_{\rm rel} \sim 2 \times 10^{-5}$ as an approximate value for the average velocity between DM particles in nearby dwarf galaxies \citep{Martinez:2010xn,McGaugh:2021tyj}.\footnote{We note that the location of the Partial-Wave Unitarity bound strongly depends on the system observed.
A search for DM annihilation within the Milky Way, for instance, would depend on a higher relative velocity, $v_{\rm rel} \sim 10^{-3}$, given the larger mass of our galaxy as compared to its satellites.
This would lower the Partial-Wave Unitarity curve shown in Fig.~\ref{fig:lim} by roughly two orders of magnitude.}
Composite states can readily evade this bound, although as shown by \cite{Griest:1989wd}, even these systems eventually hit a ``Composite Unitarity'' bound, which requires $\langle \sigma v \rangle \leq 4\pi(1+M_{\chi} v_{\rm rel} R)^2/(M_{\chi}^2 v_{\rm rel})$, which for large masses reduces to the result in Eq.~\eqref{eq:unitsc}.
We show this result for different values of $R$ in Fig.~\ref{fig:lim}, and note that for $M_{\chi} \ll R^{-1}$, these results reduce to the point-like unitarity limit.

\vspace{0.3in}

\section{Data reduction} \label{sec:data}

\subsection{VERITAS-like instrument}\label{sec:veritas}
VERITAS is an array of four imaging atmospheric Cherenkov telescopes located in Arizona, USA \citep{VERITAS}. One of the VERITAS scientific programs is to search for indirect DM signals from astrophysical objects such as dSphs and the Milky Way Galactic Center \citep{Zitzer2017}. Since it has a similar sensitivity to other IACT observatories like MAGIC and HESS \citep{Park2015, Aleksic2016, Aharonian2006}, we adopt VERITAS as representative of current-generation IACTs.

For our analysis, we take the published IRFs and observed ON and OFF region\footnote{The ON region is defined as the area centered on a target. The OFF region is one or more areas containing no known $\gamma$-ray sources, used for estimating the isotropic-diffuse background rate. } counts from \cite{dm_veritas}. The size of ON-region was 0.03 deg$^2$, and the OFF-region was defined by the crescent background method \citep{Zitzer2013}. The relative exposure time between ON and OFF regions ($\alpha$) was 0.131. From 92.0 hrs of Segue 1 observations, the number of observed events from the ON ($N_{\rm on}$) and OFF regions ($N_{\rm off}$) was 15895 and 120826, respectively. We introduce a reference instrument, denoted ``VERITAS-like,'' whose observables are limited to total $N_{\rm on}$, total $N_{\rm off}$, and $\alpha$ (see App.~\ref{sec:check} for the comparison between VERITAS and VERITAS-like constraints on the DM annihilation cross section). In addition, we scale down $N_{\rm on}$ and $N_{\rm off}$ values to a nominal observation time of 50 hours. 

\vspace{0.1in}

\subsection{CTA}\label{sec:cta}

CTA is the next-generation ground-based IACT array, which is expected to have about 10 times better point-source sensitivity when compared with the current IACT observatories, in addition to a broader sensitive energy range, stretching from 20 GeV to 300 TeV, and two to five times better energy and angular resolution \citep{Bernlohr2013}. 
The observatory will be made up of two arrays, providing full-sky coverage: one in the northern hemisphere (CTA-North; La Palma in Spain) and the other in the southern hemisphere (CTA-South; Atacama Desert in Chile). CTA will be equipped with tens of telescopes. In this study, we consider the CTA-North array, from which our target, Segue 1, can be observed. CTA will broaden our understanding of the extreme Universe, including the nature of DM \citep{CTA}, and will be able to probe long-predicted, but so-far untested candidates like Higgsino DM~\citep{Rinchiuso:2020skh}.

The CTA IRFs and background distributions as a function of energy, as well as official analysis tools,\footnote{{\tt Gammapy}, \url{https://gammapy.org/}} are publicly available \citep{CTA_IRFs, Deil2017}.
We assume the alpha configuration (\textit{prod5 v0.1}). In the alpha configuration, the CTA-North array consists of 4 Large-Sized Telescopes (LSTs) and 9 Medium-Sized Telescopes (MSTs).\footnote{\url{https://www.cta-observatory.org/science/ctao-performance/}} To compare with the VERITAS-like instrument, we use the same observation conditions; the size of the ON-region is set to 0.03 deg$^2$ with $\alpha$ of 0.131.

\vspace{0.1in}

\subsection{HAWC-like instrument}\label{sec:hawc}
HAWC, located at Sierra Negra, Mexico, is a $\gamma$-ray and cosmic-ray observatory. The instrument is constituted of 300 water tanks. Each tank contains about $1.9\times10^5$ L of water with four photomultiplier tubes (PMTs). After applying $\gamma$/hadron separation cuts, observed $\gamma$-ray events are divided into analysis bins ($\mathcal{B}_{\rm hit}$) based on the fraction of the number of PMT hits. HAWC observes two-thirds of the sky on a daily basis and has found many previously undetected VHE sources \citep{Albert2020}. In addition, they have studied 15 dSphs within its field of view to search for DM annihilation and decay signatures \citep{dm_hawc,HAWC:2017udy}.

The IRFs and observed background spectrum for Segue 1 are not publicly available, so we introduce a ``HAWC-like'' reference instrument based on reasonable assumptions. A dataset including 507 days of observations of the Crab Nebula is publicly available \citep{Abeysekara2017},\footnote{\url{https://data.hawc-observatory.org/datasets/crab_data/index.php}} and the declination angle (Dec.) of the Crab Nebula is not significantly different from that of Segue 1 ($\Delta$Dec.~$\approx$ 6 degrees). Since Dec.~is expected to be one of the key factors determining the shape of the IRFs and background rate, we assume that the background rate and IRFs should be similar for observations of Segue 1 and the Crab Nebula (see App.~\ref{sec:check} for the comparison between HAWC and HAWC-like constraints on the DM annihilation cross section). With the help of the Multi-Mission Maximum Likelihood framework \citep[{\tt 3ML}; ][]{Vianello2015}, we acquire the IRFs and background rate for each $\mathcal{B}_{\rm hit}$ (total of 9 bins) as used in \cite{Abeysekara2017}. We set the radius of an ON region to 0.2 degrees, and the background is calculated from a circular region with a 3-degree radius around the Crab Nebula, providing $\alpha$ of 0.04/9 ($\sim$ 0.004).

\vspace{0.3in}

\section{Analysis methods}\label{sec:method}

\subsection{Ingredients for estimating UHDM signal}

\begin{figure*}[t!]
    \centering
    \subfigure{\includegraphics[width=0.45\linewidth]{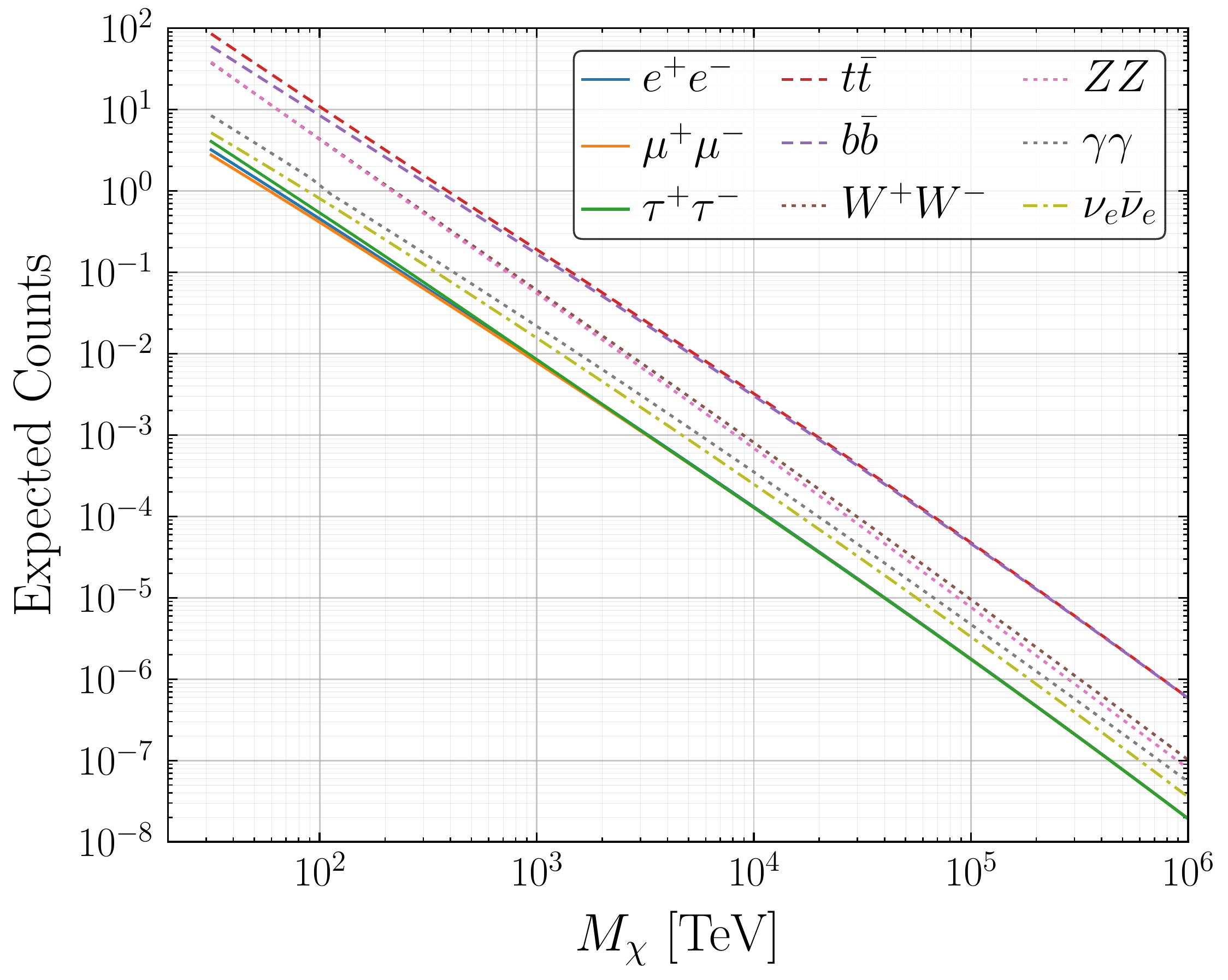}}\hspace{0.5cm}
    \subfigure{\includegraphics[width=0.45\linewidth]{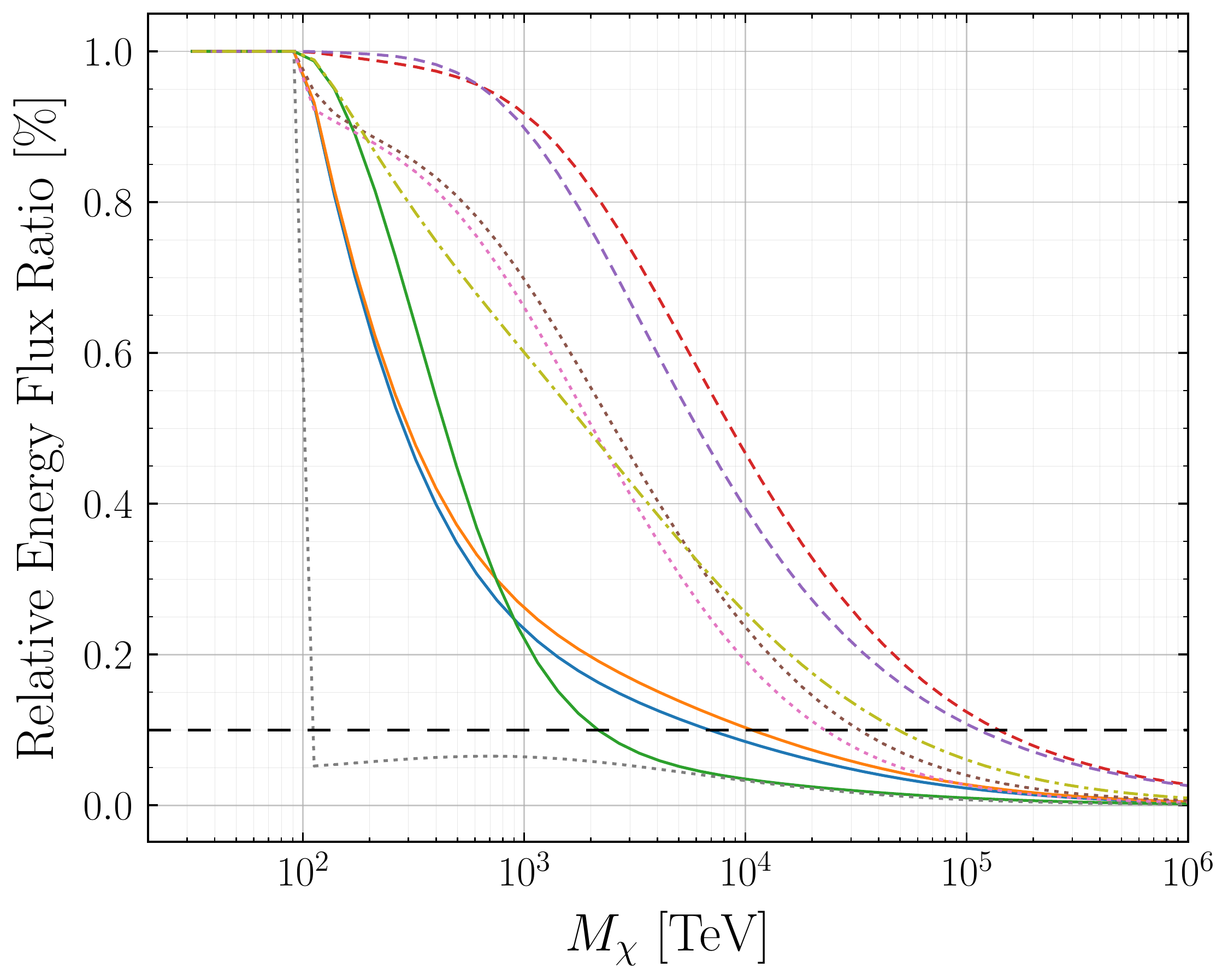}}
    \caption{The number of expected $\gamma$-ray events (left) and relative ratio between observable and total $\gamma$-ray energy flux (right).
    The expected counts are computed assuming an effective area of $10^{10}$ cm$^2$, 50 hours of exposure time, a $J$-factor of 10$^{18}$ GeV$^2$/cm$^5 \,\cdot\,$sr, and that $\langle \sigma v \rangle = 10^{-23}$ cm$^3$/s.
    The observable energy flux is defined as the integrated $\gamma$-ray energy flux up to 100 TeV, and for reference in the black dashed curve we show a value of 10\%. The portion of the observable UHDM signal from $M_{\chi} >$ 100 TeV decreases progressively as $M_{\chi}$ increases. 
    The various line styles refer to the classes of annihilation channel: charged leptons (solid), quarks (dashed), gauge bosons (dotted), and $\nu_{e}\bar{\nu}_{e}$ (dashed-dotted).}
    \label{fig:ratio}
\end{figure*}

To compute the $\gamma$-ray annihilation flux at the Earth, given in Eq.~\eqref{eq:dm_flux}, we need two ingredients: the photon spectrum for each DM annihilation channel and the DM density profile of the selected target, Segue 1. 
As stated, we use the {\tt HDMSpectrum} \citep{HDM} to calculate the expected DM signal because it provides an accurate spectrum for the full mass range we consider.
The annihilation of UHDM produces $\gamma$-rays of energies equal to or less than $M_{\chi}$.
We compute the fraction of the produced energy flux ($F \propto \int E \frac{dN}{dE} dE$) that is observable and the number of expected $\gamma$-ray events ($N \propto \int \frac{dN}{dE} dE$); i.e., the energy flux and $\gamma$-ray counts distribution within the energy band of the current and future VHE $\gamma$-ray observatories ($E \leq$ 100 TeV). 
In this work, we consider nine annihilation channels: three charged leptons ($e^{+}e^{-}$, $\mu^{+}\mu^{-}$, and $\tau^{+}\tau^{-}$), two heavy quarks ($t\bar{t}$ and $b\bar{b}$), three gauge bosons ($W^{+}W^{-}$, $ZZ$, and $\gamma\gamma$), and one neutrino ($\nu_e \bar{\nu}_e$). 

For the DM density profile, we take a generalized version of the Navarro–Frenk–White (NFW) profile, which is a function of five parameters \citep{Hernquist1990, Zhao1996, GS2015},
\begin{equation} \label{eq:dm_profile}
    \rho(r) = \frac{\rho_{s}}{(r/r_s)^{\gamma}[1+(r/r_s)^{\alpha}]^{(\beta-\gamma)/\alpha}},
\end{equation}
where the choice of ($\alpha$, $\beta$, $\gamma$) = (1, 3, 1) recovers the original NFW profile \citep{NFW1997}, and $r_s$ is the scale radius of the DM halo.
The so-called $J$-factor is defined as the integral of the squared DM density along the los within a region of interest (roi), 
\begin{equation}
    J = \int_{\rm roi}d\Omega \int_{\rm los}dl\, \rho^2(l\hat{n}).
\end{equation}
The set of five NFW parameters ($\alpha$, $\beta$, $\gamma$, $\rho_s$, and $r_s$) is obtained by fitting the observed kinematic data of the dSphs. Limited data produces large uncertainties in estimates of the $J$-factor, which propagates as a systematic uncertainty when estimating the DM cross section (see Sec.~\ref{sec:discussion}). In a thorough study, \cite{GS2015} obtained a number of the parameter sets that adequately describe the data. Among more than 6000 sets for Segue 1, we take one that approximates the median of the $J$-factor (see Table~\ref{tab:nfw}).

Fig.~\ref{fig:ratio} shows the expected number of $\gamma$-ray photons under the conditions stated below (left panel) and the ratio of observable energy flux to total energy flux (right panel) for the nine annihilation channels. For the expected counts distribution, we assume that the effective area is $10^{10}$ cm$^2$, the exposure time is 50 hours, the $J$-factor is 10$^{18}$ GeV$^2$/cm$^5 \,\cdot\,$sr, and the DM cross section is 10$^{-23}$ cm$^3$/s.
This result implies that the current and future observatories, whose sensitive energy ranges extend to 100 TeV, can observe a large portion of the produced $\gamma$-rays and/or energy flux from the UHDM annihilation, up to $M_{\chi}$ of a few PeV.
For the $\gamma \gamma$ channel, the majority of the energy remains in the sharp spectral feature at $E_\gamma \sim M_{\chi}$, and so the energy flux ratio sharply drops once the mass is above 100 TeV and the continuum component becomes dominant. This sharp decrease is not clearly visible in the expected count level because the emission at $E_\gamma \sim M_{\chi}$ produces only about 10\% of the total counts in the high-mass regime.

\begin{table}[h!]
	\centering 
	\begin{tabular}{c c c c c c c c}
    \hline\hline
    $\rho_s$ & $r_s$ & $\alpha$ & $\beta$ & $\gamma$ & $\theta_{\rm max}$ & $J(\theta_{\rm max})$ \\ 
    $[$ \(M_\odot\)$/{\rm pc}^3$ ] & [ pc ] & & & & [ deg ] & [ GeV$^2$/cm$^5 \,\cdot\,$sr ] \\ \hline
    $5.1\times10^{-3}$& $2.2\times10^4$ & 1.48 & 8.04 & 0.83 & 0.35 & $2.5\times10^{19}$\\
    \hline\hline
    \end{tabular}
    \caption{The selected parameter set of the generalized NFW profile for Segue 1. The maximum angular distance, $\theta_{\rm max}$ is given by the location of the furthest member star, which is an estimate of the size of Segue 1.}\label{tab:nfw}
\end{table}

\vspace{0.1in}

\subsection{Projected sensitivity curves}\label{sec:excess}
To explore the feasibility of detection, we compare expected $\gamma$-ray counts from UHDM self-annihilation to background counts. The number of expected signal counts ($N_s$) is obtained by forward-folding Eq.~\eqref{eq:dm_flux} with IRFs, 
\begin{equation}\label{eq:dm_signal}
N_s = \int d\Omega\, dE'\, dE\, \frac{dF(E', \hat{n})}{dE'\, d\Omega} R(E, \Omega|E', \Omega'),
\end{equation}
where unprimed and primed quantities represent observed (strictly speaking, reconstructed) and true quantities, respectively. The function $R(E, \Omega|E', \Omega')$, refers to an IRF consisting of three sub-functions: effective area, energy bias, and point spread function. Assuming that the number of ON region events is $N_{\rm on} = N_s+\alpha N_{\rm off}$, we calculate the significance of the UHDM signal by using the so-called Li \& Ma significance \citep[$\mathcal{S}$;][]{Li1983},
\begin{equation}
    \mathcal{S} = \sqrt{2} \left\{ N_{\rm on} \ln \left[ \frac{1+\alpha}{\alpha} \left( \frac{N_{\rm on}}{N_{\rm on}+N_{\rm off}} \right) \right] + N_{\rm off} \ln \left[ (1+\alpha) \left( \frac{N_{\rm off}}{N_{\rm on}+N_{\rm off}} \right) \right] \right\}^{1/2}.
\end{equation}
Finally, for each annihilation channel, we find a set of values of $M_{\chi}$ and $\langle\sigma v\rangle$ for which $\mathcal{S}$ = 5\,$\sigma$.

\subsection{Expected upper limit curves}\label{sec:uls}
To estimate an UL on the UHDM annihilation cross section for a given $M_\chi$, we perform a maximum likelihood estimation (MLE). Since we cannot access the energy distribution of background events for the VERITAS-like instrument, we use a simple likelihood analysis using the total $N_{\rm on}$ and $N_{\rm off}$ counts, $\mathcal{L}(\langle\sigma v\rangle; b|D)$, constructed from two Poisson distributions,
\begin{equation}
\begin{aligned}
    \mathcal{L} &= \mathcal{P}_{\rm pois} \left( N_s + \alpha b; N_{\rm on} \right) \times \mathcal{P}_{\rm pois}(b; N_{\rm off})\\
    &= \frac{ \left( N_s + \alpha b \right) ^{N_{\rm on} } e^{-(N_s+\alpha b)}}{N_{\rm on}!}\frac{b^{N_{\rm off}}e^{-b}}{N_{\rm off}!}, 
\end{aligned}
\end{equation}
where the nuisance parameter $b$ represents the expected background rate. This likelihood function is expected to be less sensitive compared to a full likelihood function incorporating event-wise energy information, especially at high masses, as it does not utilize any features present in the DM spectrum; see \cite{Aleksic2012} for full discussion of this hindrance. For CTA and the HAWC-like instrument, we perform a binned likelihood analysis, 
\begin{equation}
    \mathcal{L} = \sum_i \frac{ \left( N_{s, i} + \alpha b \right) ^{N_{{\rm on}, i} } e^{-(N_{s, i}+\alpha b)}}{N_{{\rm on}, i}!}\frac{b^{N_{{\rm off}, i}}e^{-b}}{N_{{\rm off}, i}!}.
\end{equation}

We calculate an expected UL with the assumption that an ON region does not contain any signal from UHDM self-annihilation but only Poisson fluctuation around $\alpha \times N_{\rm off}$; i.e., we can randomly sample $N_{\rm on}$ from the Poisson distribution of $\alpha N_{\rm off}$. For the binned likelihood analysis, we can apply the Poisson fluctuation to each background bin to get the binned ON-region data. With the synthesized ON-region data, we perform MLE analysis and calculate an UL on the DM cross section for a given $M_{\chi}$. Throughout this paper, UL refers to the one-sided 95\% confidence interval, which is obtained from the profile likelihood ($\Delta \ln\mathcal{L} = 1.35$). We repeat the process of calculating an expected limit to get the median or the containment band for the 95\% UL.

\vspace{0.3in}

\section{Results} \label{sec:result}
\begin{figure*}[t!]
    \centering
    \subfigure{\includegraphics[width=0.32\linewidth]{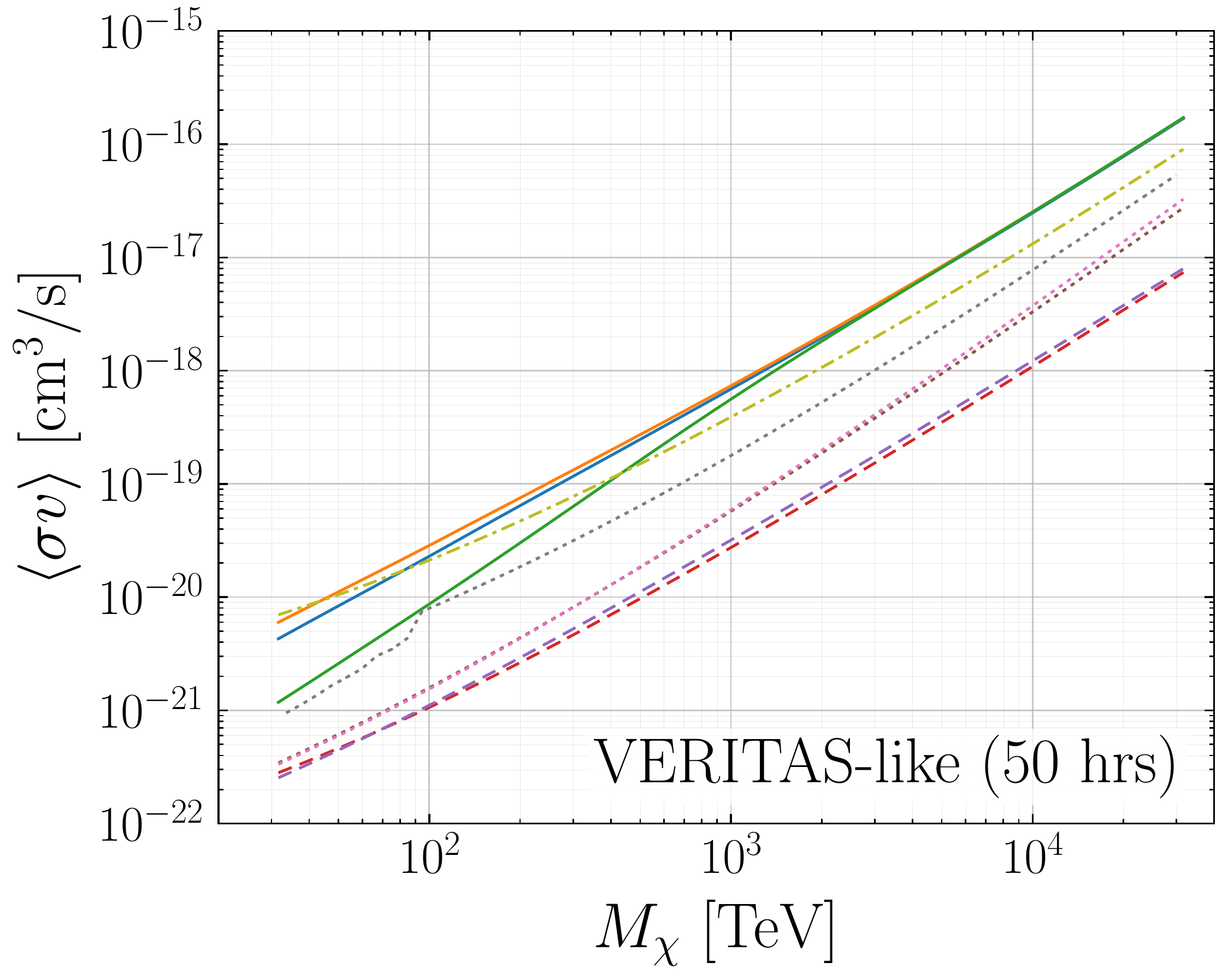}}
    \subfigure{\includegraphics[width=0.32\linewidth]{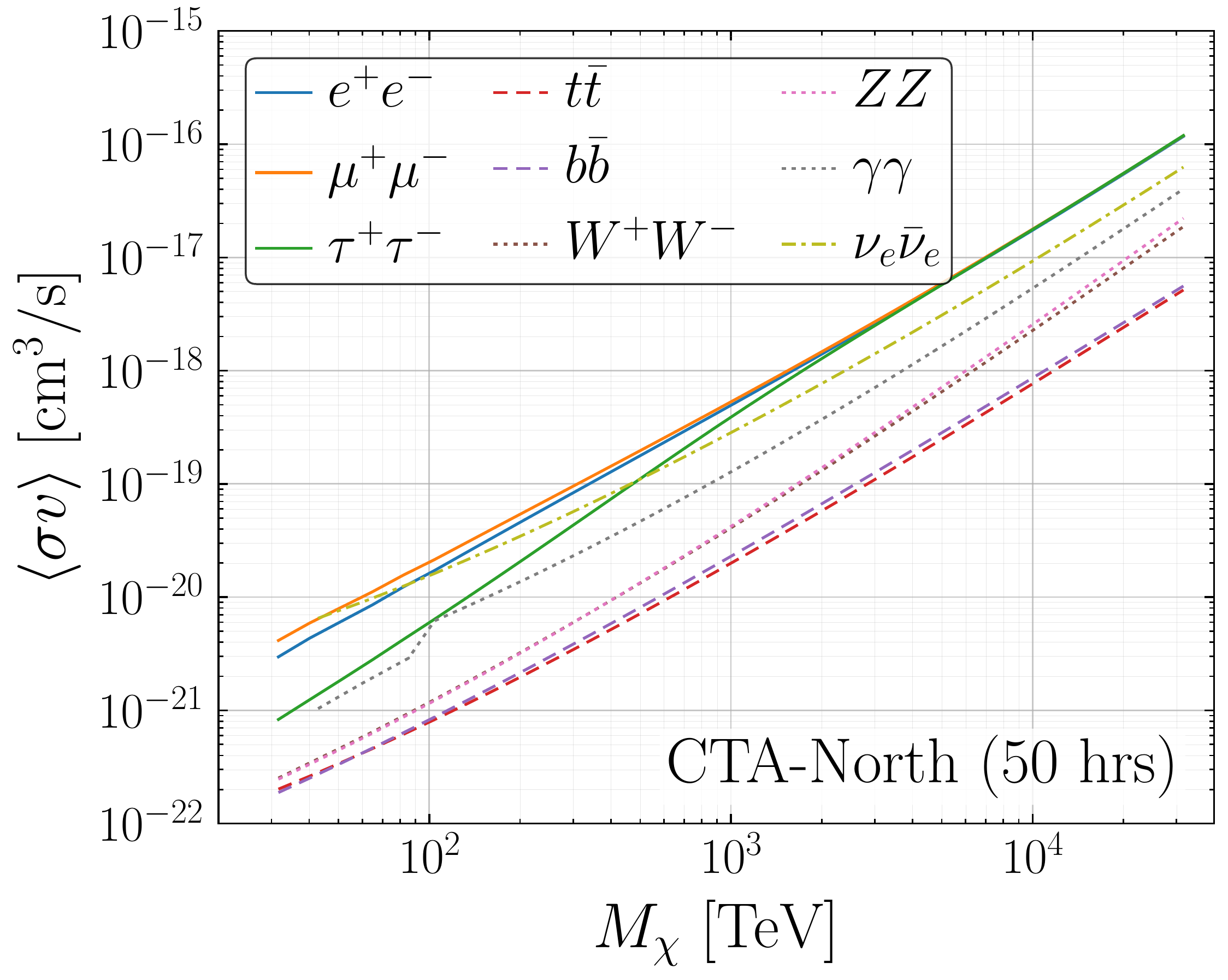}}
    \subfigure{\includegraphics[width=0.32\linewidth]{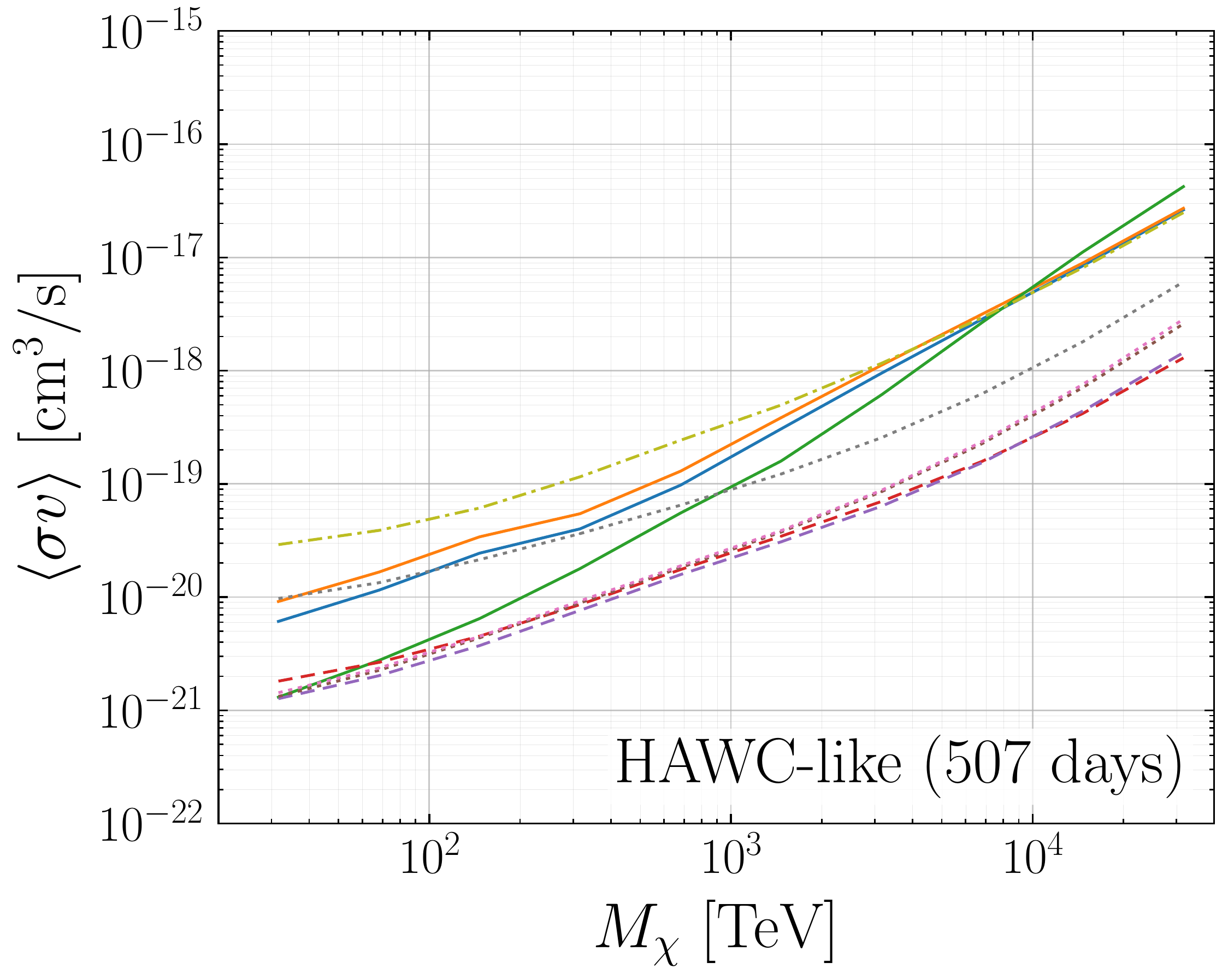}}
    \caption{Sensitivity curves for the nine UHDM annihilation channels for a VERITAS-like instrument (50 hrs; left panel), CTA-North (50 hrs; middle panel), and a HAWC-like instrument (507 days; right panel). Each curve corresponds to a set of parameters ($M_{\chi}$ and $\langle\sigma v\rangle$), producing a 5\,$\sigma$ signal excess (Sec.~\ref{sec:excess}). Line styles are as in Fig.~\ref{fig:ratio}.}
    \label{fig:sensitivity}
\end{figure*}

Here, we present two sets of analysis results: sensitivity curves and expected ULs, as functions of the UHDM particle mass. Since above a few tens of PeV the energy flux ratio for all annihilation channels is less than 10\% (Fig.~\ref{fig:ratio}), we perform the analyses for UHDM masses from 30 TeV up to 30 PeV. Note that all of the following results are based on assumed exposure times of 50 hours for the VERITAS-like instrument and CTA-North, and 507 days for the HAWC-like instrument.

Figure~\ref{fig:sensitivity} shows the sensitivity curves for nine UHDM annihilation channels ($e^{+}e^{-}$, $\mu^{+}\mu^{-}$, $t\bar{t}$, $b\bar{b}$, $W^{+}W^{-}$, $ZZ$, $\gamma\gamma$\footnote{Note that for the $\gamma\gamma$ channel, we use a different mass binning so that the lower bound of the sensitivity and upper limit curves is different from those from the other channels. This choice is based on the fact that the delta component in the $\gamma\gamma$ annihilation can be fully addressed only when the mass binning matches the binning of the energy bias matrix ($M_\chi = E_\gamma$).}, and $\nu_e \bar{\nu}_e$) with VERITAS-like (50 hrs; left panel), CTA-North (50 hrs; middle panel), and HAWC-like (507 days; right panel) instruments. Considering the annihilation of an UHDM particle with $M_{\chi}$ of 1 PeV via the $\tau^{+}\tau^{-}$ channel, a HAWC-like instrument is likely to reach $\mathcal{S}$ of 5\,$\sigma$ with the smallest cross section; specifically, a VERITAS-like instrument is expected to detect UHDM for a cross section of $\sim 5\times10^{-19}~{\rm cm}^3/{\rm s}$, CTA-North for $\sim 4\times10^{-19}~{\rm cm}^3/{\rm s}$, and a HAWC-like instrument for $\sim 1\times10^{-19}~{\rm cm}^3/{\rm s}$. However, this sensitivity depends on the annihilation channel and the UHDM mass, not to mention the exposure time. For example, for $M_{\chi}$ of 100 TeV, CTA-North shows, in general, a better sensitivity compared to the other instruments. For the $\gamma \gamma$ channel, a discontinuity in the sensitivity lines can be seen because, as explained earlier, the line-like contribution ($E_\gamma \sim M_\chi$) falls outside the sensitive energy range.

\begin{figure*}[t!]
    \centering
    \subfigure{\includegraphics[width=0.32\linewidth]{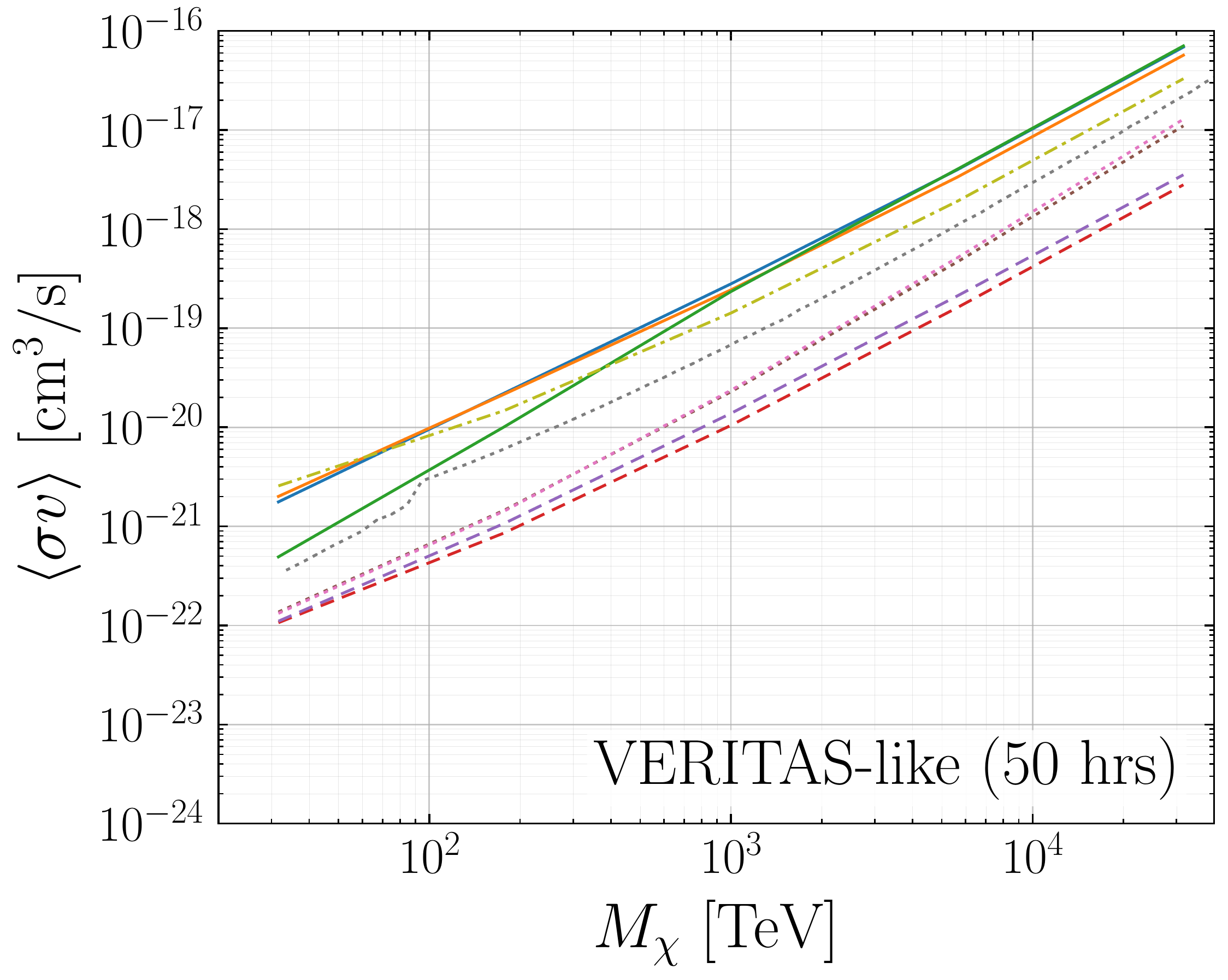}}
    \subfigure{\includegraphics[width=0.32\linewidth]{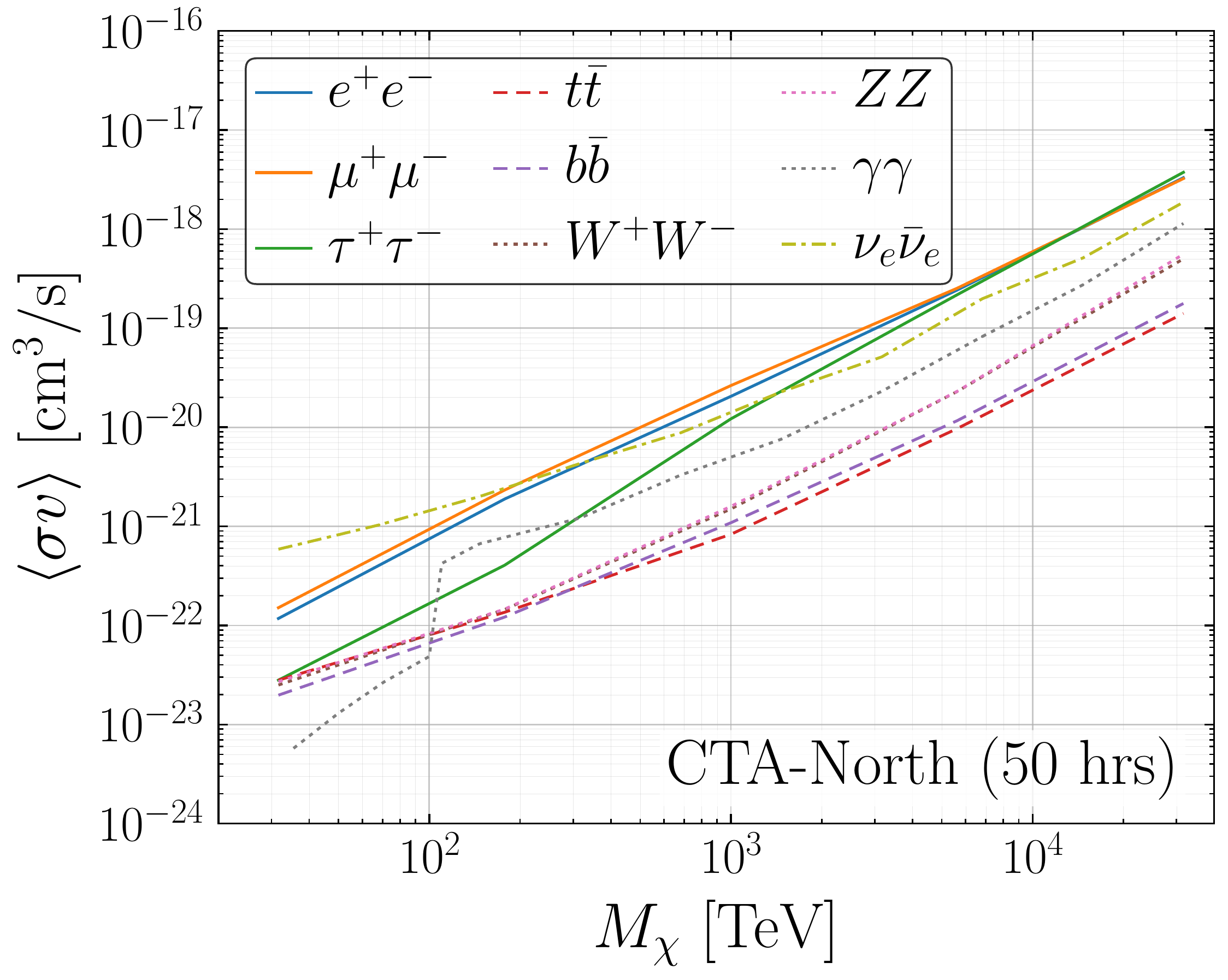}}
    \subfigure{\includegraphics[width=0.32\linewidth]{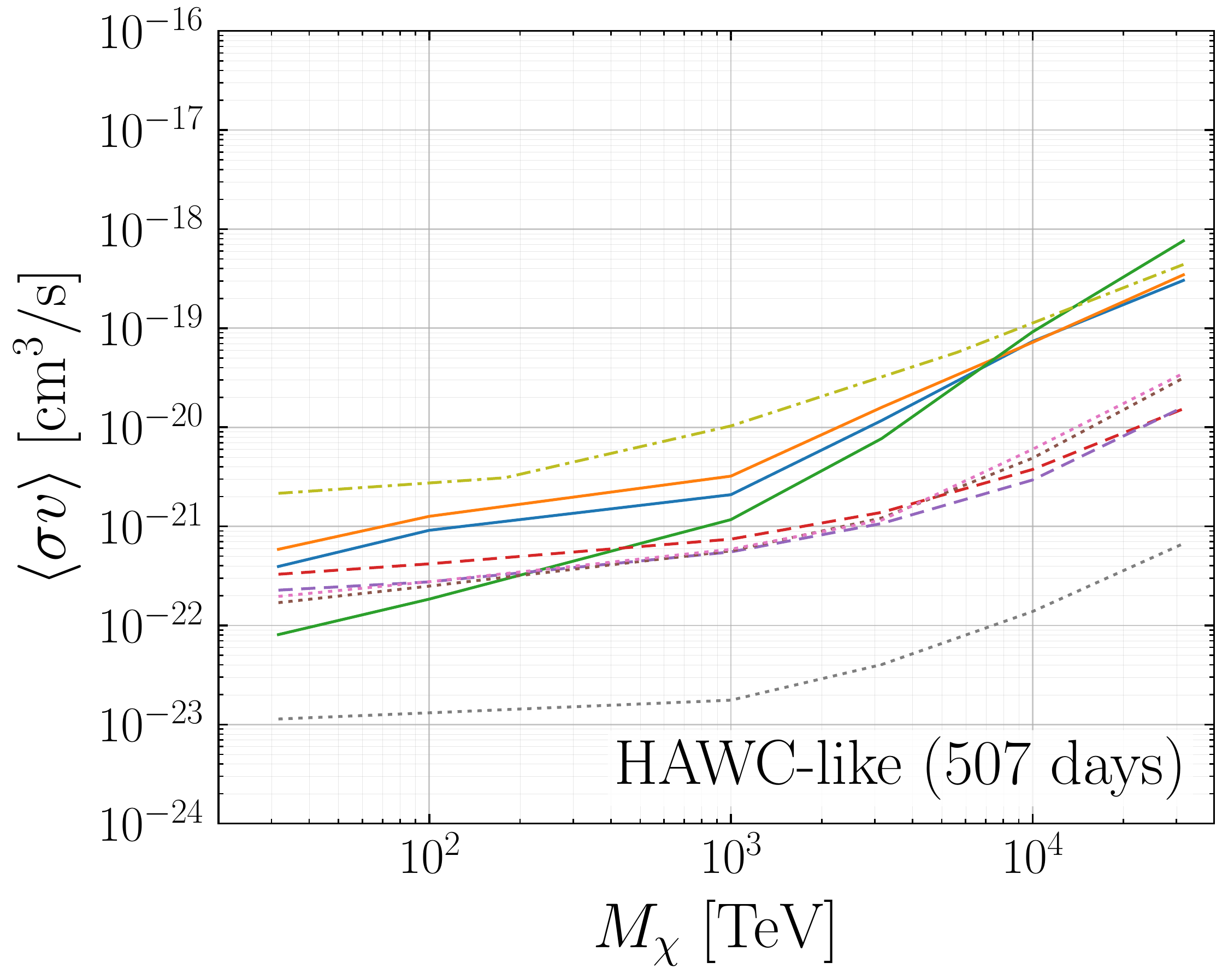}}
    \caption{Expected 95\,\% UL curves of UHDM cross section for the nine UHDM annihilation channels, obtained for a VERITAS-like instrument (50 hrs; left panel), CTA-North (50 hrs; middle panel), and a HAWC-like instrument (507 days; right panel). An expected UL is the median of 100 realizations obtained from the profile likelihood, assuming no signal excess in an ON region (Sec.~\ref{sec:uls}). Again, the line styles follow Fig.~\ref{fig:ratio}.}
    \label{fig:uls}
\end{figure*}

Next, we estimate the ULs on the UHDM annihilation cross section as a function of UHDM particle mass for the same annihilation channels for the three instruments (Fig.~\ref{fig:uls}). The curves represent the median value from 100 realizations generated at each mass. With the assumed observation conditions (e.g., livetime), CTA-North shows the most constraining ULs at lower masses ($M_{\chi} < 1$ PeV), whereas a HAWC-like instrument provides more stringent ULs at higher masses. Note that the UL on the DM cross section is expected to decrease as we increase the exposure time, $\langle\sigma v\rangle_{\rm UL} \propto 1/\sqrt{t}$. As expected from the relative sensitivity between VERITAS and CTA-North, the UL curves from CTA-North are about 10 times lower than those from a VERITAS-like instrument. 

In the case of the $\gamma\gamma$ annihilation channel, a discontinuity in the UL curve is again observed at 100 TeV, most strongly for the CTA-North. In contrast to the VERITAS-like instrument, it is possible for the CTA-North instrument to perform the full binned likelihood analysis by comparing the signal and background energy distributions, which lowers the UL curve (see App.~\ref{sec:check}). Note that in the case of the $\gamma\gamma$ annihilation channel, the two distributions differ clearly compared to those of other channels. In the case of the HAWC-like instrument, the energy dispersion matrix for the highest energy bin is relatively broad, which smooths out the discontinuity.

\vspace{0.3in}

\section{Discussion of statistical and systematic uncertainties} \label{sec:discussion}

Here we briefly discuss the impact of statistical and systematic uncertainties on the presented UL curves. For these studies, we consider a single annihilation channel ($t\bar{t}$) for simplicity, although the results are representative of what we expect for the additional channels.

\begin{figure}[t!]
    \centering
    \subfigure{\includegraphics[width=0.45\linewidth]{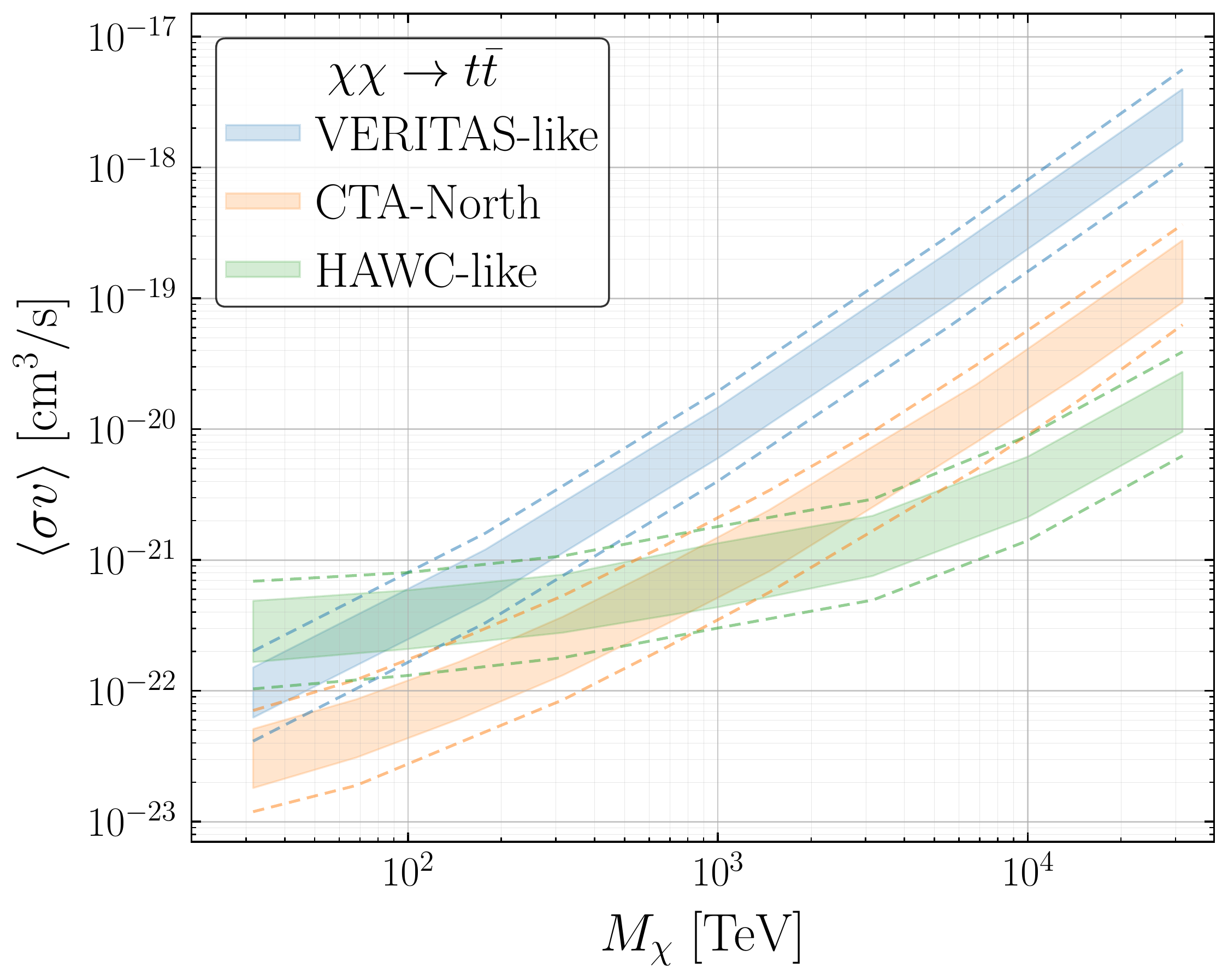}}\hspace{0.5cm}
    \subfigure{\includegraphics[width=0.45\linewidth]{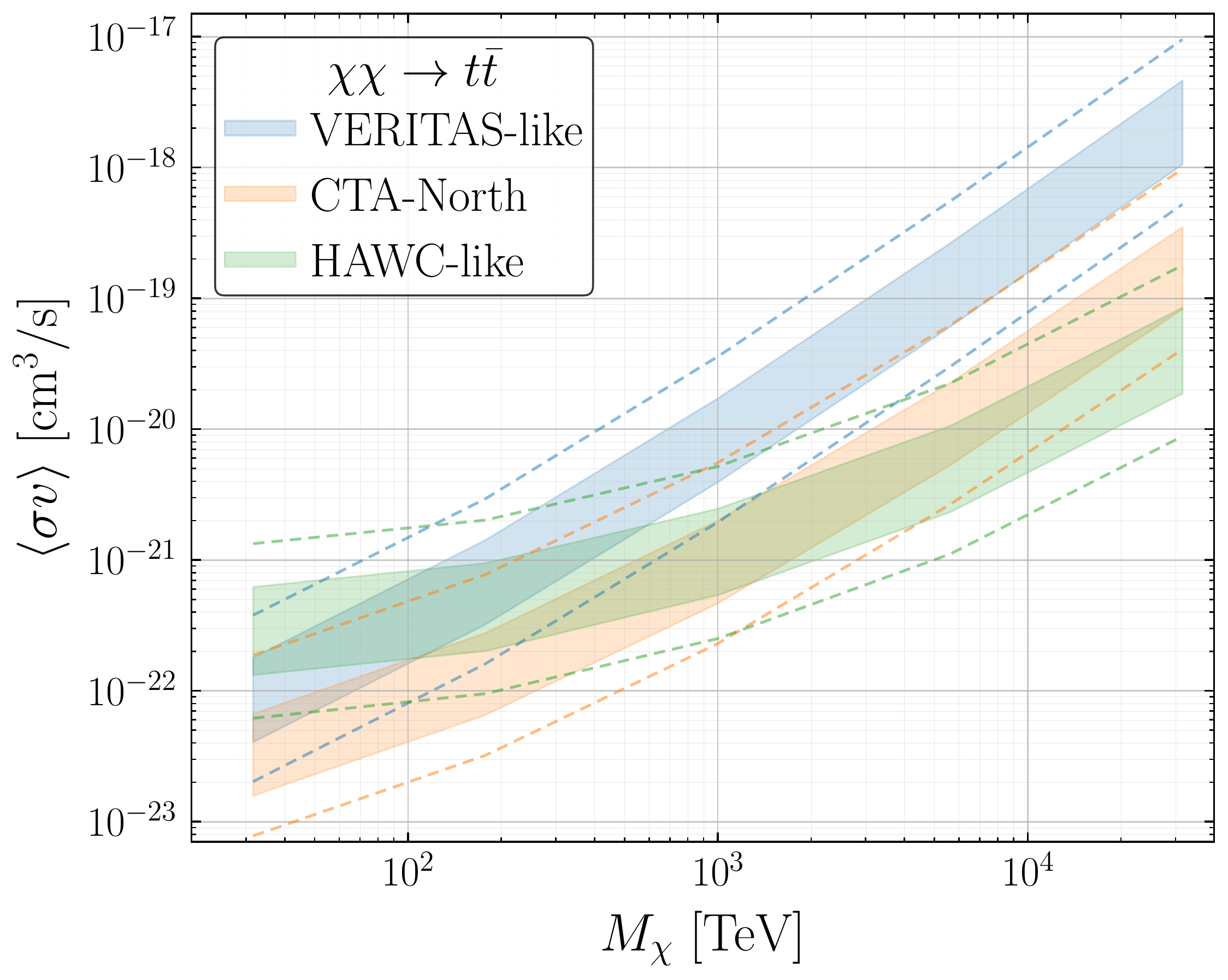}}
    \caption{{\it Left:} Statistical uncertainty on the expected 95\% limits. Each uncertainty band is obtained from 10$^{4}$ realizations for the $t\bar{t}$ annihilation channel.
    {\it Right:} Systematic uncertainty on the same expected limits, resulting from uncertainties in the $J$-factor estimation. Each uncertainty band is obtained from 10$^{4}$ realizations for the $t\bar{t}$ annihilation channel.
    In both figures, shaded region refers to 68\% containment, and dashed lines are 95\% containment.}
    \label{fig:statsys_err}
\end{figure}

Due to the Poisson fluctuation in the observed counts, statistical uncertainty is inevitable. For this study, we compute the 68\% containment band of expected UL curves for a large number of MC realizations (10,000), using the method described in Sec.~\ref{sec:uls}.
Figure~\ref{fig:statsys_err} shows the statistical uncertainty band for 68\% (shaded region) and 95\% (dashed lines) containment. This figure implies that the Poisson fluctuation can result in 45--55\% statistical uncertainty (at the 1$\sigma$ level) across all masses for the three instruments: VERITAS-like ($\sim$45\%), CTA-North ($\sim$53\%), and HAWC-like ($\sim$54\%). 

A major systematic uncertainty, beyond that inherent in IRFs, is the present uncertainty in the DM density profile assumed for Segue 1.
A DM density profile estimated from insufficient and possibly inaccurate kinematic observations will inevitably have a large uncertainty.
Also, it depends on assumptions and approximations made in the modeling---for instance the assumption of a NFW profile with exact spherical symmetry---can also lead to systematic uncertainties.
In addition, the stellar sample selection when fitting the DM density profile affects the $J$-factor significantly, such that any ambiguity in the sample selection, possibly due to contamination from foreground stars or stellar streams, can overestimate the $J$-factor.
The magnitudes of the systematic uncertainties are different from dSph to dSph, and depend on the definition of the DM density profile. For further discussion on this uncertainty, see \cite{Bonnivard2015a, Bonnivard2015b}. 

As mentioned earlier, \cite{GS2015} provide more than 6000 viable parameter sets for Segue 1, and we compute $10^{4}$ expected UL curves by randomly sampling the parameter set. In this work, we use the parameter sets to estimate the systematic uncertainty on an expected UL curve due to uncertainty on the $J$-profile. Note that in this study, we do not include the Poisson fluctuation of the simulated ON region counts; i.e., $N_{{\rm on}, i}$ is equal to $\alpha N_{{\rm off}, i}$. Finally, we take ULs corresponding to the 68\% and 95\% containment for each mass (Fig.~\ref{fig:statsys_err}). This figure implies that, for Segue 1, the $J$-factor can increase or decrease an UL curve by a factor of 2 (1\,$\sigma$ level) across all masses, regardless of instrumental properties, at a level to the statistical uncertainties seen in Fig.~\ref{fig:statsys_err}. Note that \cite{Bonnivard2016} claimed that $J$-factor may be overestimated by about two orders of magnitude due to the stellar sample selection bias. However, the accurate prediction of the Segue 1 $J$-profile is beyond the scope of this paper.

\vspace{0.3in}

\section{Summary and Outlook}\label{sec:summary}

In this work, we have explored the potential of current and future $\gamma$-ray observatories to extend the search for DM beyond the unitarity bound.
Our results allow one to determine whether discovery of an UHDM candidate of a given mass and annihilation cross section is within reach. Furthermore, we provide an estimate of the constraints that can be derived on the UHDM annihilation cross section by current and future $\gamma$-ray observatories, assuming a non-detection.

Returning to Fig.~\ref{fig:lim}, we can place our obtained limits in the context of theoretical constraints on the allowed annihilation cross section of UHDM. All instruments considered can probe realistic cross sections for composite UHDM particles whose annihilation respects partial-wave unitary. For the given exposure times (50 hours for CTA-North and a VERITAS-like instrument, and 507 days for a HAWC-like instrument), CTA-North is projected to provide the most constraining limits, probing scales down to $R = (10~{\rm GeV})^{-1} $ for UHDM with a mass around 300~TeV. At higher masses, above 1~PeV, HAWC-like limits become the most constraining, reaching scales around $R = (1~{\rm GeV})^{-1}$ at 10~PeV. The VERITAS-like limits, while less constraining, are worse than those of CTA-North or a HAWC-like instrument by less than or equal to an order of magnitude for the entire mass range (with a slight advantage over the HAWC-like instrument at masses below 100~TeV).

This work draws attention to the exploration of DM beyond the conventional parameter range. The results we have derived are indicative, using reasonable assumptions about the data and IRFs for current-generation instruments, as well as realistic exposure times for current and future instruments. We hope that this work illustrates the interest and feasibility of searches for UHDM with the current-generation $\gamma$-ray instruments, and the value of considering such searches for future observatories. The phase space that can be probed, in terms of DM particle mass and annihilation cross section, is a relevant one for models predicting composite UHDM. This parameter space is currently unconstrained, but could be probed with archival datasets from current-generation $\gamma$-ray instruments, including HAWC, VERITAS, and other IACTs.

\vspace{0.1in}
\begin{acknowledgments}

{\it Acknowledgments.}
Our work benefited from discussions with Michael Geller, Diego Redigolo, and Juri Smirnov. 
We would like to thank Alex Geringer-Sameth, Savvas M. Koushiappas, and Matthew Walker for providing the parameter sets for the $J$-factors. 
This research has made use of the CTA instrument response functions provided by the CTA Consortium and Observatory, see https://www.cta-observatory.org/science/cta-performance/ (version prod5 v0.1; [citation]) for more details.
D. Tak and E. Pueschel acknowledge the Young Investigators Program of the Helmholtz Association, and additionally acknowledge support from DESY, a member of the Helmholtz Association HGF. M. Baumgart is supported by the DOE (HEP) Award DE-SC0019470.
%

\end{acknowledgments}

\bibliographystyle{aasjournal}
\bibliography{references}

\appendix

\section{Comparing reference and real instruments}\label{sec:check}

In this appendix, we perform a consistency check showing that our two reference instruments (VERITAS-like and HAWC-like) can qualitatively reproduce the published results from VERITAS \citep[92.0 hrs for Segue 1;][]{dm_veritas} and HAWC \citep[507 days for Segue 1;][]{dm_hawc}. In particular, we compute an expected UL band and then compare it with the corresponding published UL curves. At the outset, we emphasize that given the ingredients for estimating the DM annihilation signal (e.g., DM density profile) as well as the method for computing ULs differ from the publications, we do not expect complete consistency.

\begin{figure}[h!]
    \centering
    \subfigure{\includegraphics[width=0.45\linewidth]{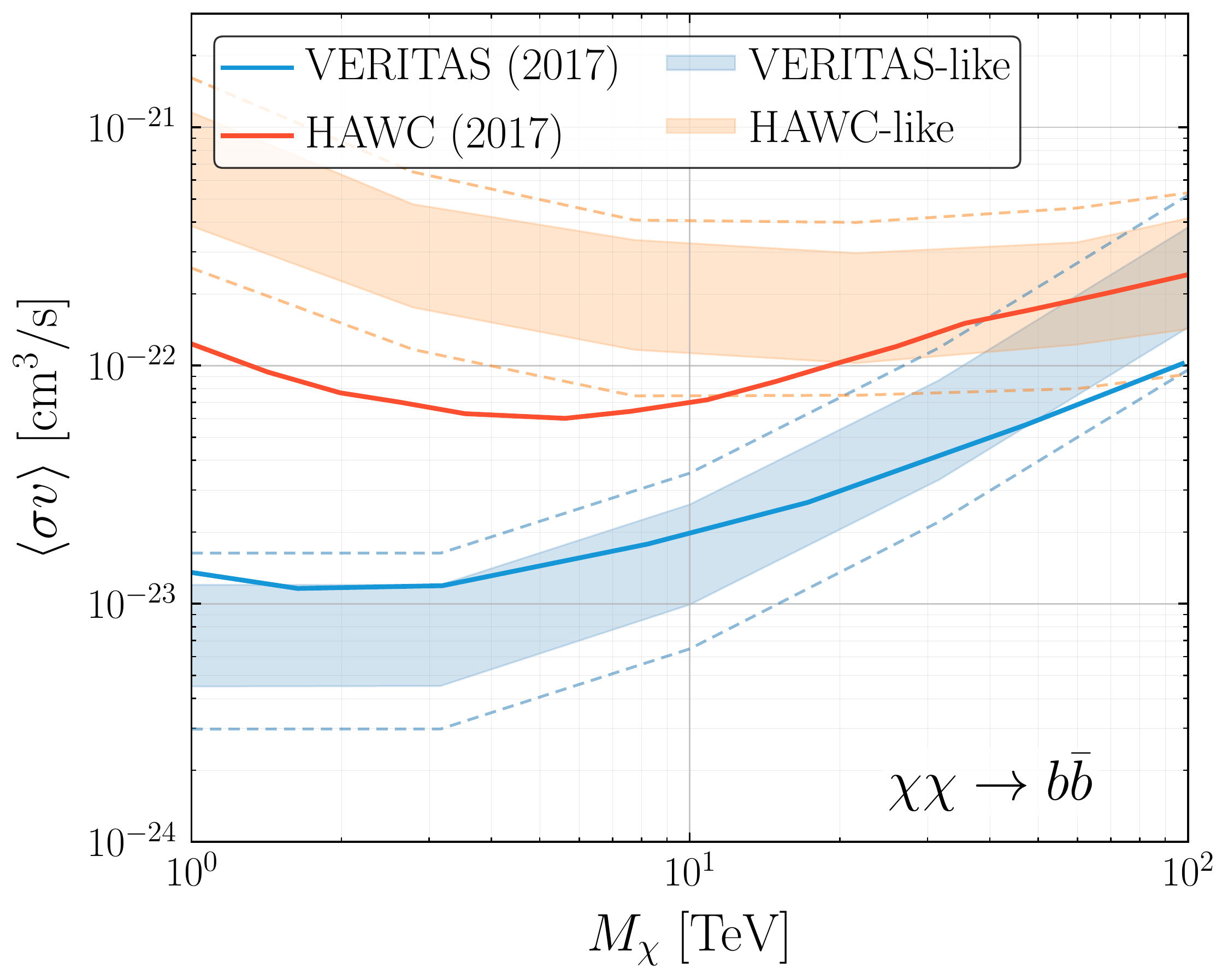}}\hspace{0.5cm}
    \subfigure{\includegraphics[width=0.45\linewidth]{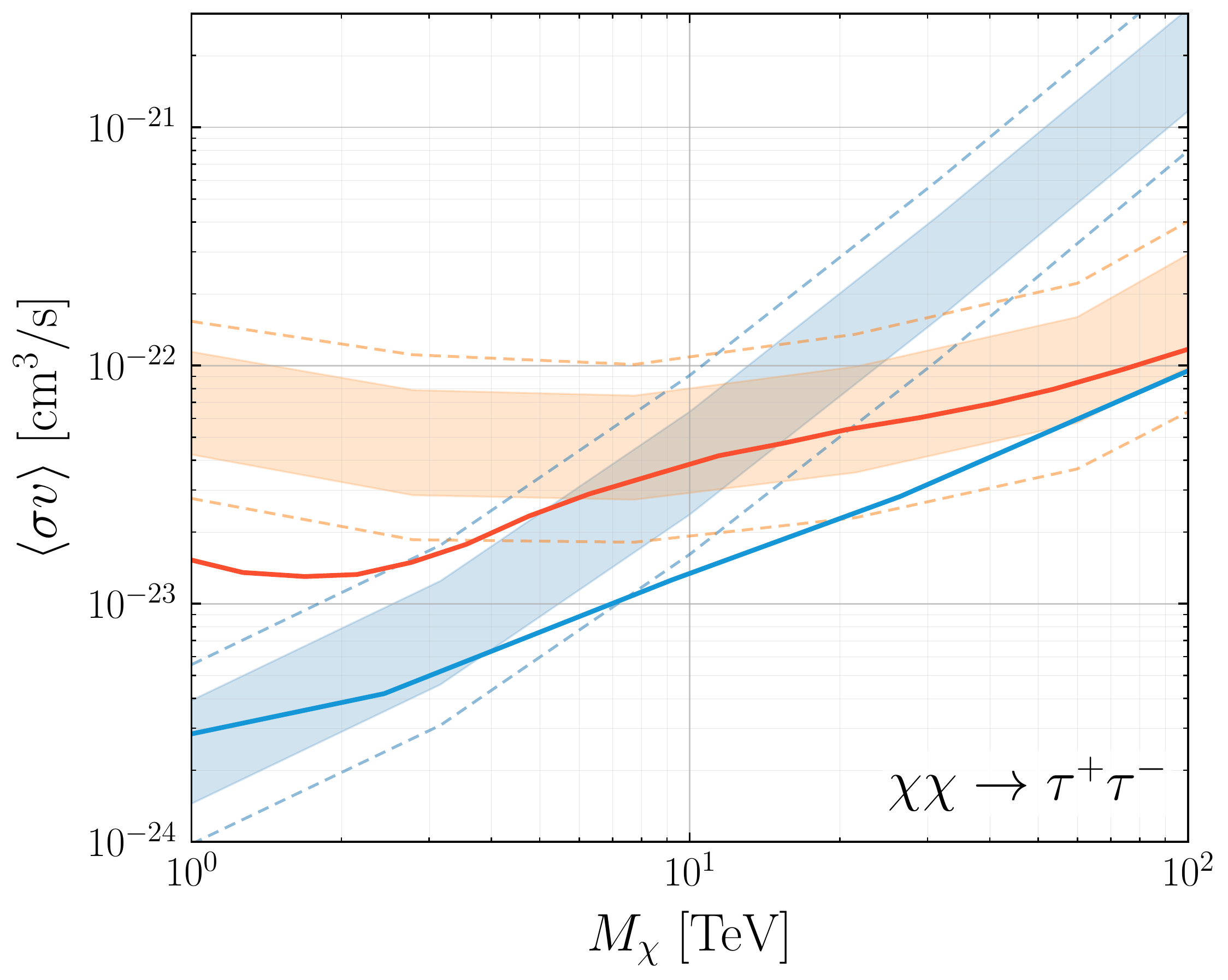}}
    \caption{Comparison between the expected 95\% limits for our reference instruments and the equivalent published curves. Both curves and bands show the 95\% confidence bands on the DM annihilation cross section for the $b\bar{b}$ (left) and $\tau^{+}\tau^{-}$ (right) annihilation channels, obtained from Segue 1. Each expected UL band is from MC simulations, where a shaded region (upper and lower dashed lines) corresponds to the 68\% (95\%) containment band. A solid line represents a published UL curve, where the VERITAS and HAWC results are from \cite{dm_veritas} and \cite{dm_hawc}, respectively.}
    \label{fig:sanityCheck}
\end{figure}

As described in Sec.~\ref{sec:uls}, each expected UL curve can be obtained from a single MC simulation, and an UL band is based on 300 realizations. 
The consistency check is performed for the $b\bar{b}$ and $\tau^{+}\tau^{-}$ annihilation channels because both \cite{dm_veritas} and \cite{dm_hawc} provide those UL curves for Segue 1.
In Fig.~\ref{fig:sanityCheck} we show the comparison of the expected UL bands and the published UL curves for the two instruments, VERITAS-like and HAWC-like.
In the case of the VERITAS-like instrument, the published UL curve and the expected UL band are consistent at lower masses ($M_{\chi} \sim 1-10~{\rm TeV}$). However, as the DM mass increases, the two UL curves deviate. As mentioned earlier and discussed in \cite{Aleksic2012}, the deviation can be easily explained by the fact that the likelihood function used for the VERITAS-like instrument lacks sensitivity at high masses. It can be resolved if we perform the full binned/unbinned likelihood analysis with the actual dataset. However, moving to the full likelihood is beyond the scope of this paper.
Meanwhile, the result for the HAWC-like instrument shows greater consistency with the published result in the high-mass regime ($M_{\chi}\gtrsim 10$ TeV). The discrepancy at lower masses arises mostly from the assumptions we adopted on the HAWC-like instrument, described in Sec.~\ref{sec:hawc}; in particular, we have assumed the HAWC observation of Segue 1 exactly matches that of the Crab Nebula, for which we used the available observed background values and IRFs.

\end{document}